\pdfoutput=1
\documentclass[superscriptaddress,twocolumn,aps,preprintnumbers,amsmath,amssymb,nofootinbib]{revtex4}

\usepackage{amsmath}
\usepackage{amssymb}
\usepackage{amsfonts}
\usepackage{graphicx}
\usepackage[dvipsnames]{xcolor}
\usepackage{xfrac}
\usepackage{comment}
\usepackage{pifont}
\usepackage{physics}
\usepackage{fourier}
\usepackage{hyperref}
\usepackage{bm}
\usepackage{enumitem}
\usepackage{makecell}
\usepackage{siunitx} 
\usepackage{enumitem}

\usepackage{makecell}
\usepackage{siunitx} 
\usepackage{tikz} 

\usepackage{float}
\usepackage{subcaption} 

\definecolor{rossoferrari}{HTML}{D9073D}
\definecolor{mediumblue}{HTML}{0000CD}
\definecolor{forestgreen}{HTML}{228B22}
\definecolor{desy_blue}{HTML}{009EE2}
\definecolor{desy_orange}{HTML}{FD8800}
\definecolor{light_pink}{rgb}{1,0.4,0.4}
\definecolor{light_blue}{rgb}{0.284602,0.317763,0.963947}
\hypersetup{setpagesize=false,bookmarksnumbered=true,bookmarksopen=true,%
colorlinks=true,linkcolor=light_blue,urlcolor=rossoferrari,citecolor=rossoferrari,linktocpage=false}

\newcommand{\diag}{\operatorname{diag}}

\begin{document}


\preprint{MITP-25-086}

\title{Freeze-in leptogenesis without the need for right-handed neutrino oscillations}

\author{Martin A.\ Mojahed}
\email{martin.mojahed@roma1.infn.it}
\affiliation{INFN Sezione di Roma, Piazzale Aldo Moro 2, I-00185 Rome, Italy}

\author{Sascha Weber}
\email{wesascha@uni-mainz.de}
\affiliation{PRISMA$^+$ Cluster of Excellence \& Mainz Institute for Theoretical Physics, FB 08 - Physics, Mathematics and Computer Science, Johannes Gutenberg-Universität Mainz, Staudingerweg 9, 55099 Mainz, Germany}


\begin{abstract}
We present an experimentally testable leptogenesis mechanism based on the standard type-I seesaw model that successfully operates at right-handed-neutrino (RHN) masses around the GeV scale. The mechanism takes place in a cosmological background with an asymmetry between right-handed electrons and left-handed positrons generated at high temperatures, and does not require oscillations between RHNs or any CP violation in the RHN sector. In contrast to standard leptogenesis via freeze-in, our mechanism works even in the presence of a single RHN around the GeV scale. The mechanism is illustrated for the minimal type-I seesaw with two RHNs, where we show that successful baryogenesis via leptogenesis is viable in large regions of parameter space even without a small mass splitting between the RHNs.

\end{abstract}


\date{\today}
\maketitle


\noindent\textbf{Introduction}\,---\, The baryon asymmetry of the Universe (BAU) constitutes one of the most compelling indications for physics beyond the Standard Model (SM). An attractive extension of the SM capable of addressing the BAU puzzle through baryogenesis via leptogenesis (LG)~\cite{Fukugita:1986hr} is the so-called type-I seesaw model~\cite{Minkowski:1977sc,Yanagida:1979as,Yanagida:1980xy,Gell-Mann:1979vob,Mohapatra:1979ia}, which extends the SM field content with at least two RHNs and thus provides a direct link between the BAU and the origin of tiny masses for the active neutrinos~\cite{Buchmuller:2005eh,Chun:2017spz,Bodeker:2020ghk}. In its standard realization, LG can be categorized into two distinct dynamical regimes. The first, freeze-out LG (FOLG), relies on CP-violating, out-of-equilibrium decays of heavy RHNs and constitutes the traditional formulation of the mechanism~\cite{Fukugita:1986hr,Luty:1992un,Covi:1996wh,Pilaftsis:1997jf,Barbieri:1999ma,Pilaftsis:2003gt,Giudice:2003jh,Buchmuller:2004nz,Pilaftsis:2005rv,Davidson:2008bu}. The second, freeze-in LG (FILG), instead proceeds predominantly through CP-violating oscillations among RHNs in the thermal plasma of the early Universe~\cite{Akhmedov:1998qx,Asaka:2005pn}. 

FOLG provides the dominant production mechanism for RHN masses from the GUT scale down to $\mathcal{O}(100)$ GeV. For RHNs roughly heavier than the so-called Davidson-Ibarra bound~\cite{Davidson:2002qv,Buchmuller:2002rq}, $M_N\geq 10^{9-10}$ GeV, FOLG may be realized with a hierarchical RHN mass spectrum. For lighter RHNs, successful baryogenesis via FOLG typically necessitates either the inclusion of flavour-dependent dynamics in the lepton sector~\cite{Moffat:2018wke} or a resonant enhancement of the CP asymmetry in the RHN sector~\cite{Covi:1996wh,Pilaftsis:1997jf,Pilaftsis:2003gt,Pilaftsis:2005rv}, the latter requiring a mass spectrum with very small splittings between at least two RHNs. For RHN masses of the order $\mathcal{O}\left(100\right)$ GeV, the allowed parameter space connects smoothly to the freeze-in regime~\cite{Klaric:2020phc,Klaric:2021cpi}, which provides the dominant production channel for even lighter RHNs. Also successful baryogenesis via FILG typically requires small mass splittings between RHNs, in particular in the minimal type-I seesaw realization with only two heavy RHNs, see e.g.~\cite{Klaric:2020phc}.

The mechanisms described so far rely on all three of Sakharov's conditions for baryogenesis~\cite{Sakharov:1967dj} being satisfied in the RHN sector with sufficient efficiency and magnitude to generate the BAU. Crucially, these standard LG realizations depend on sufficient CP violation in the RHN sector, which enforces very tight constraints on mass splittings and Yukawa couplings if RHNs have masses well below the Davidson-Ibarra bound. Recently, it has been demonstrated that the stringent constraints associated with successful FOLG can be substantially alleviated if the mechanism operates in a cosmological background in which one or more of the numerous global charges that are conserved by SM interactions at high temperatures possess nonvanishing values at the time of LG. This framework, dubbed wash-in LG (WILG)~\cite{Domcke:2020quw,WILGRelated}, does not rely on CP violation in the RHN sector. Instead, RHNs are only needed to act as a source of $B-L$ violation that can drive the thermal plasma to an attractor solution featuring nonzero $B-L$ charge, even if $B-L$ has been conserved in all preceding processes throughout the history of the universe. Hence, WILG should be viewed as an effective description that requires novel CP-violating dynamics at high temperatures, which generates a nonminimal cosmological background upon which the RHNs act, see Refs.~\cite{Domcke:2022kfs,Schmitz:2023pfy,Mukaida:2024eqi,Mojahed:2024yus} for some realizations. In~\cite{Mojahed:2025vgf}, it was shown that WILG via freeze-out of RHNs with a hierarchical mass spectrum can successfully account for the BAU as long as the lightest RHN is heavier than roughly $10$ TeV, which is many orders of magnitude below the Davidson-Ibarra bound.

The purpose of this \textit{letter} is to demonstrate that, without imposing any additional assumptions beyond those in Ref.~\cite{Mojahed:2025vgf}, WILG can successfully operate down to the big bang nucleosynthesis (BBN) bound on RHN masses $M_N \gtrsim\SI{0.1}{\giga\electronvolt}$ (see e.g. Ref.~\cite{Boyarsky:2020dzc} and references therein) if it proceeds via freeze-in of RHNs rather than freeze-out. In contrast to standard FILG, the WILG mechanism does not rely on RHN oscillations or mixing, and can therefore be realized even if the RHN mass spectrum features only a single light RHN. As we will show, large regions of the neutrino-parameter space leading to successful generation of the BAU through wash-in freeze-in LG (WIFI-LG) are within reach of planned terrestrial experiments, making it a testable LG scenario.

The remainder of this paper is organized as follows. We first present the quantum kinetic equations (QKEs) governing WIFI-LG, followed by a simple parametric estimate of the resulting baryon asymmetry to build basic intuition. We then provide numerical solutions to the QKEs and identify regions of parameter space in which WIFI-LG can account for the BAU in the minimal type-I seesaw extension of the SM. Finally, we discuss potential avenues for future work and conclude. Some additional results and comparisons between standard FILG and WIFI-LG are collected in the appendix. 

\noindent\textbf{WIFI-LG}\,---\, 
The equations governing the evolution of RHNs and SM particles in WIFI-LG are similar to those employed in standard FILG, which we review in the following. The standard QKEs~\cite{Akhmedov:1998qx,Asaka:2005pn,Ghiglieri:2017gjz,Abada:2018oly,Granelli:2023vcm} for FILG with an arbitrary number of RHNs takes the following form in the density matrix formalism~\cite{Sigl:1993ctk},
\begin{widetext}
\begin{align}
    \frac{T_{\text{EW}}}{M_0}\dv{R_N}{z} &=-i\frac{z}{T_{\text{EW}}} \left[\expval{H},R_N\right] - \frac{1}{2}\frac{\expval{\gamma_N^{(0)}}}{T}\left\{F^{\dagger}F,R_N-1\right\} +\frac{\expval{\gamma_N^{(1)}}}{T}F^{\dagger}\mu F \nonumber - \frac{1}{2} \frac{\expval{\gamma_N^{(2)}}}{T}\left\{F^{\dagger}\mu F,R_N\right\} \nonumber \\
    &- \frac{z^2}{2T_{\text{EW}}^2}\frac{\expval{S_N^{(0)}}}{T}\left\{MF^{T}F^{*}M,R_N-1\right\} -\frac{z^2}{T_{\text{EW}}^2}\frac{\expval{S_N^{(1)}}}{T}MF^{T}\mu F^{*}M +\frac{z^2}{2T_{\text{EW}}^2} \frac{\expval{S_N^{(2)}}}{T}\left\{MF^{T}\mu F^{*}M,R_N\right\}, \label{eq:QKE1} \\
    \frac{2\pi^2}{9\zeta(3)}\frac{T_{\text{EW}}}{M_0}\dv{\mu_{\Delta_{\alpha}}}{z}&=-\frac{1}{2}\frac{\expval{\gamma_N^{(0)}}}{T} (FR_NF^{\dagger}-F^{*}R_{\Bar{N}}F^{T})_{\alpha\alpha} + \frac{\expval{\gamma_N^{(1)}}}{T}(FF^{\dagger})_{\alpha\alpha} \mu_{\alpha} -\frac{1}{2}\frac{\expval{\gamma_N^{(2)}}}{T} (FR_NF^{\dagger}+F^{*}R_{\Bar{N}}F^{T})_{\alpha\alpha} \mu_{\alpha} \nonumber \\
    &+\frac{z^2}{2T_{\text{EW}}^2}\frac{\expval{S_N^{(0)}}}{T} (F^{*}MR_NMF^{T}-FMR_{\Bar{N}}MF^{\dagger})_{\alpha\alpha} + \frac{z^2}{T_{\text{EW}}^2}\frac{\expval{S_N^{(1)}}}{T}(FM^2F^{\dagger})_{\alpha\alpha} \mu_{\alpha} \nonumber \\
    &-\frac{z^2}{2T_{\text{EW}}^2}\frac{\expval{S_N^{(2)}}}{T} (FMR_{\Bar{N}}MF^{\dagger}+F^{*}MR_{N}MF^{T})_{\alpha\alpha} \mu_{\alpha}. \label{eq:QKE2}
\end{align}
\end{widetext}
Here $H=\sqrt{\frac{\pi^2g_*}{90}}\frac{T^2}{M_P}$ denotes the Hubble rate for a standard radiation-dominated universe, $z=T_{\text{EW}}/T$ where $T_{\text{EW}}=131.7$ GeV~\cite{DOnofrio:2014rug} is the temperature of sphaleron decoupling, $M_N$ is the $n\times n$ mass matrix for the RHNs, $F$ is the standard $3\times n$ Yukawa coupling between the three active and $n$ sterile neutrinos, and $M_0=T^2/H\approx \SI{7.12e17}{\giga\electronvolt}$. 
The $n\times n$ matrix $R_N$ encodes the number density matrix of the RHNs, normalized to the equilibrium number density, $R_N(T)=n_N(T)/n_N^{\rm eq}(T)$,
and the equations for $R_{\Bar{N}}$ and $R_{N}$ are related via CP conjugation. The momentum-averaged effective Hamiltonian is given by $\expval{H}$~\cite{Hernandez:2016kel,Abada:2018oly} and $\expval{\gamma^{(i)}}$ $\left(\expval{s^{(i)}}\right)$~\cite{Ghiglieri:2017gjz,Hernandez:2022ivz} denote lepton-number conserving (lepton-number violating) momentum-averaged dissipation rates. Finally, the SM is described by the following chemical potentials normalized to the temperature to have zero mass dimension $\mu = \diag(\mu_{\alpha}) = \diag(\mu_{\ell_{\alpha}}+\mu_{\phi})$, where $\mu_{\ell_\alpha}$ are the flavored left-chiral lepton chemical potentials and $\mu_\phi$ is the Higgs chemical potential. The combinations $\mu_\alpha$ are related to lepton-flavor asymmetries, $\Delta_\alpha=B/3-L_\alpha$, through spectator effects.
Indeed, in the literature on FILG~\cite{Abada:2018oly}, the following linear relation $\mu_{\alpha} = -2 \sum_{\beta}C_{\alpha\beta}\mu_{\Delta_{\beta}}$ is commonly applied, where the flavor coupling matrix $C_{\alpha\beta}$ encodes spectator process in the thermal plasma~\cite{Buchmuller:2001sr,Garbrecht:2014kda,Garbrecht:2019zaa}. However, in a fully general description, the said linear relation should only be understood as a limiting case of the following affine relation
\begin{equation}\label{eq:affine_relation}
    \mu_{\alpha} = 2\left[\mu_{\alpha}^0- \sum_{\beta}C_{\alpha\beta}\mu_{\Delta_{\beta}}\right]
\end{equation}
where $\mu_\alpha^{0}$ encodes global charges (expressed in units of chemical potentials) that are approximately conserved by SM processes at high temperatures.
Both $\mu_\alpha^{0}$ and $C_{\alpha\beta}$ evolve as a function of temperature, as systematically outlined in~\cite{Antaramian:1993nt,Fong:2015vna,Domcke:2020quw,factorTwoComment}. While the fully general relation in Eq.~\eqref{eq:affine_relation} has been mentioned in the FILG literature, see e.g.~\cite{Garbrecht:2014bfa}, so far only the case $\mu_{\alpha}^0=0$ has been studied in this context. The novelty of our work lies in realizing that a nonzero value of $\mu_{\alpha}^0$, respecting both theoretical and experimental bounds, can have non-negligible impact on the predictions of low-scale LG.

For the remainder of the work, we will consider the possibility that an asymmetry between right-handed electrons and left-handed positrons, $q_e = n_e - \bar{n}_e=\mu_{e_R}T^3/6$, was generated at a temperature $T\gtrsim\SI{e6}{\giga\electronvolt}$. The size of this asymmetry is constrained from above as~\cite{Bodeker:2019ajh}, 
\begin{align}
    \label{eq:upperbound}
    \abs{\mu_{e_R}} \lesssim \num{9.6 e-4},
\end{align}
by requiring that the primordial chiral asymmetry stored in right-handed electrons must not trigger the chiral instability of the SM plasma~\cite{Joyce:1997uy,Boyarsky:2011uy,Akamatsu:2013pjd,Hirono:2015rla,Yamamoto:2016xtu,Rogachevskii:2017uyc,Kamada:2018tcs,Gurgenidze:2025lpt}. In our scenario, the affine relation Eq.~\eqref{eq:affine_relation} is given by~\cite{Domcke:2020quw}
\begin{equation}
    \mu_{\alpha}^0 =\begin{pmatrix}
        -\frac{5}{13}\\
        \frac{4}{37}\\
        \frac{4}{37}
    \end{pmatrix} \mu_{e_R}, \qquad C = \begin{pmatrix}
        \frac{6}{13}&0&0\\
        0&\frac{41}{111}&\frac{4}{111}\\
        0&\frac{4}{111}&\frac{41}{111}
    \end{pmatrix},
\end{equation}
for $T\in\left[T_e,\SI{e6}{\giga\electronvolt}\right]$, where $T_e\approx\SI{85}{\tera\electronvolt}$ denotes the equilibration temperature of the Yukawa interaction of the right-handed electrons~\cite{Bodeker:2019ajh}. The validity of the above equations can be extended from $T_e$ down to the electroweak phase transition, $T_{\rm{EW}}$, by coupling Eqs.~\eqref{eq:QKE1}-\eqref{eq:QKE2} with the evolution equation for $\mu_{e_R}$~\cite{Bodeker:2019ajh}
\begin{equation}
\label{eq:beq_eR}
\frac{d\mu_{e_R}}{dz} = - \frac{\Gamma_e}{zH}\left[\frac{711}{481}\,\mu_{e_R} + \frac{5}{13}\,\mu_{\Delta_e} - \frac{4}{37}\left(\mu_{\Delta_\mu} + \mu_{\Delta_\tau}\right) \right] \,,
\end{equation}
where $\Gamma_e$ is the equilibration rate of right-handed electrons, see e.g.~\cite{Mojahed:2025vgf} for details about $\Gamma_e$. Finally, the resulting baryon asymmetry generated via WIFI-LG is obtained by evaluating $\mu_{\Delta_\alpha}$ at $T_{\rm{EW}}$, 
\begin{align}
    \label{eq:YB}
    Y_B=c_{\rm{sph}}\sum_\alpha\frac{15 \mu_{\Delta_\alpha}}{2\pi^2g_{*,s}}\Bigg |_{T_{\text{EW}}},
\end{align}
where $g_{*,s}=106.75$ is the number of entropic degrees of freedom and $c_{\rm{sph}}$ is the sphaleron conversion factor~\cite{Harvey:1990qw,Laine:1999wv}.


\noindent\textbf{Simple estimate}\,---\, In general, the coupled system of evolution equations, Eqs.~\eqref{eq:QKE1}-\eqref{eq:QKE2} and Eq.~\eqref{eq:beq_eR}, must be solved numerically. Before doing so, let us first present a rough parametric estimate to illustrate how various physical quantities entering the problem may affect the final asymmetry. For simplicity, the following estimate is restricted to scenarios with slow equilibration of RHNs. We set all lepton-number-violating rates, i.e.~$\expval{s^{(i)}}$, in Eqs.~\eqref{eq:QKE1}-\eqref{eq:QKE2} to zero~\cite{LNVcomment}, and ignore oscillations between RHNs~\cite{Osccomment}. We further approximate scattering rates as follows, $\expval{\gamma^{(i)}}=c_i T$, where $c_i$ are temperature-independent constants~\cite{Hernandez:2022ivz,Granelli:2023vcm}. Finally, let us approximate the evolution of $\mu_{e_R}$ by treating it as a constant above $T_e$ and instantaneously equilibrated at $T_e$, i.e., 
\begin{align}
    \mu_{e_R} = \mu_{e_R}^0\theta(T-T_e)+\mu_{e_R}^{\rm eq}\theta(T_e-T),
\end{align}
where $\mu_{e_R}^{\rm eq}$ is defined by the vanishing of the right-hand side of Eq.~\eqref{eq:beq_eR}. For minimal initial conditions, $R_N(T=\SI{e6}{\giga\electronvolt}) \approx \mu_{\alpha}(T=\SI{e6}{\giga\electronvolt}) \approx 0$, the leading contribution to Eq.~\eqref{eq:QKE2} is of the following form, 
\begin{equation}
    \label{eq:8}
    \frac{2\pi^2}{9\zeta(3)}\frac{T_{\text{EW}}}{M_0}\dv{\mu_{\Delta_{\alpha}}}{z}\approx \frac{\expval{\gamma_N^{(1)}}}{T}(FF^{\dagger})_{\alpha\alpha} \mu_{\alpha}.
\end{equation}
The solution can be written as,
\begin{align}
    \label{eq:approximation}
    \mu_{\Delta_{\alpha}}(z_e) &\approx \frac{9\zeta(3)}{\pi^2} \frac{z_e}{z_1} (\Hat{F}\Hat{F}^{\dagger})_{\alpha\alpha} \mu_{\alpha}^{0},\\
    \label{eq:rough2}
    \mu_{\Delta_{\alpha}}(z_{\text{EW}}) &\approx \left[\exp(-\frac{9\zeta(3)}{\pi^2}\frac{A}{z_1})\right]_{\alpha\beta} \mu_{\Delta_{\beta}}(z_e),
\end{align}
where $\frac{1}{z_1} = c_1\frac{M_0}{T_{\text{EW}}} \tr(FF^{\dagger})$, and we have defined
\begin{align}
     \Hat{F} = \frac{F}{\sqrt{\tr(FF^{\dagger})}}, \quad A_{\alpha\beta} = (\Hat{F}\Hat{F}^{\dagger})_{\alpha\alpha}C_{\alpha\beta}.
\end{align}
Physically, Eq.~\eqref{eq:approximation} represents the lepton–flavor asymmetry washed in by the RHNs during the epoch in which $\mu_{e_R}$ is still sizable, i.e. for $T \gtrsim T_e$. In contrast, for temperatures in the range $T_{\rm EW} \lesssim T \lesssim T_e$, $\mu_{e_R}$ is rapidly depleted, and the RHNs begin to wash out lepton-flavor asymmetries in the same way as in conventional LG. Eq.~\eqref{eq:rough2} determines the residual asymmetry at sphaleron decoupling.

To understand how our scenario may evade devastating wash-out effects, note that the right-hand side of Eq.~\eqref{eq:8} is proportional to $\mu_{e_R}^0$ during wash-in and proportional to $\mu_\Delta$ during wash-out. A hierarchy between $\mu_{\Delta_\alpha}(z_e)$ and $\mu_\alpha^0$ can accommodate wash-in while preventing large wash-out effects. A second point is that wash-in and wash-out are sensitive to different combinations of entries in the Yukawa matrices, implying that there can be textures that enhance wash-in while suppressing wash-out, and vice versa.

\begin{figure*}%
    \centering
    {{\includegraphics[height=0.199\textheight]{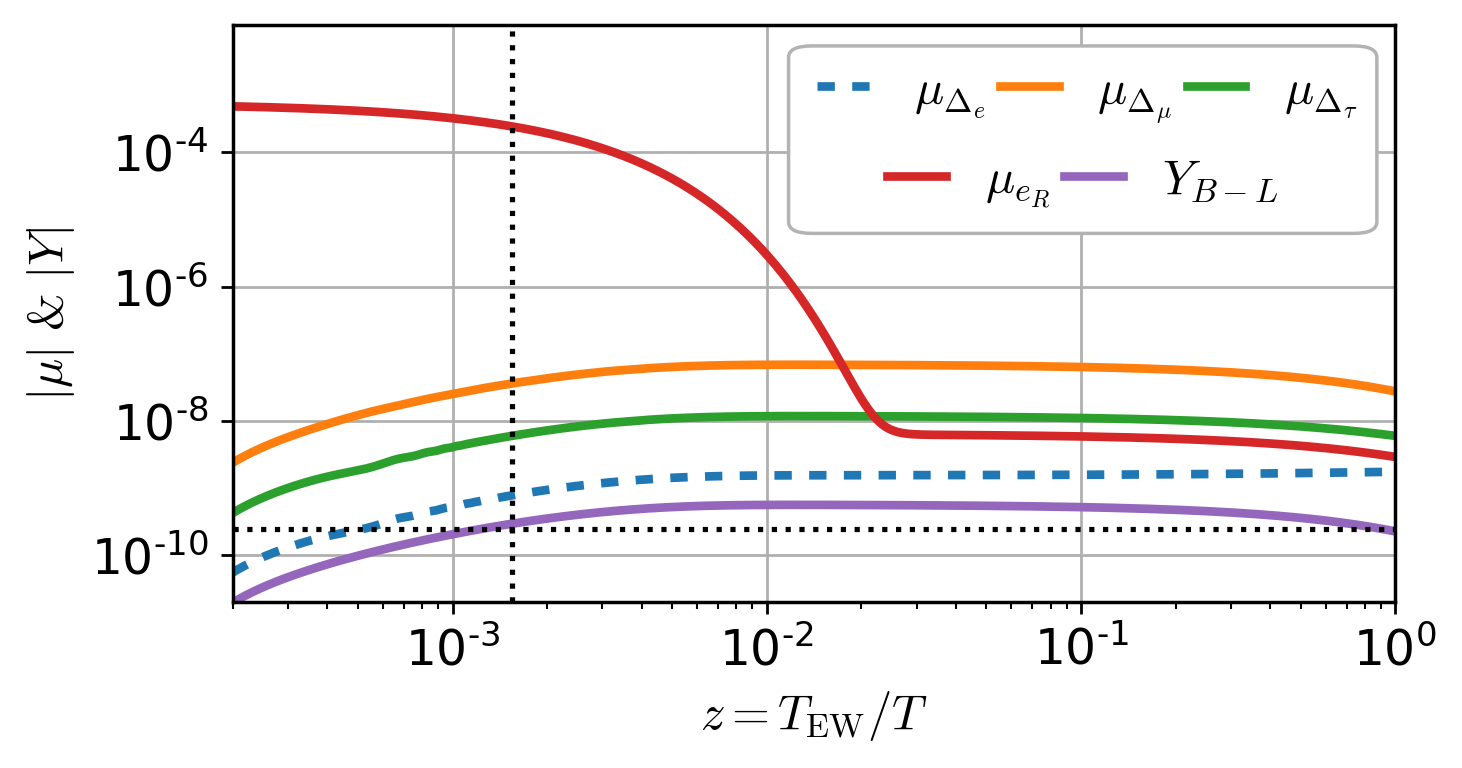}}}%
    \quad
    {{\includegraphics[height=0.199\textheight]{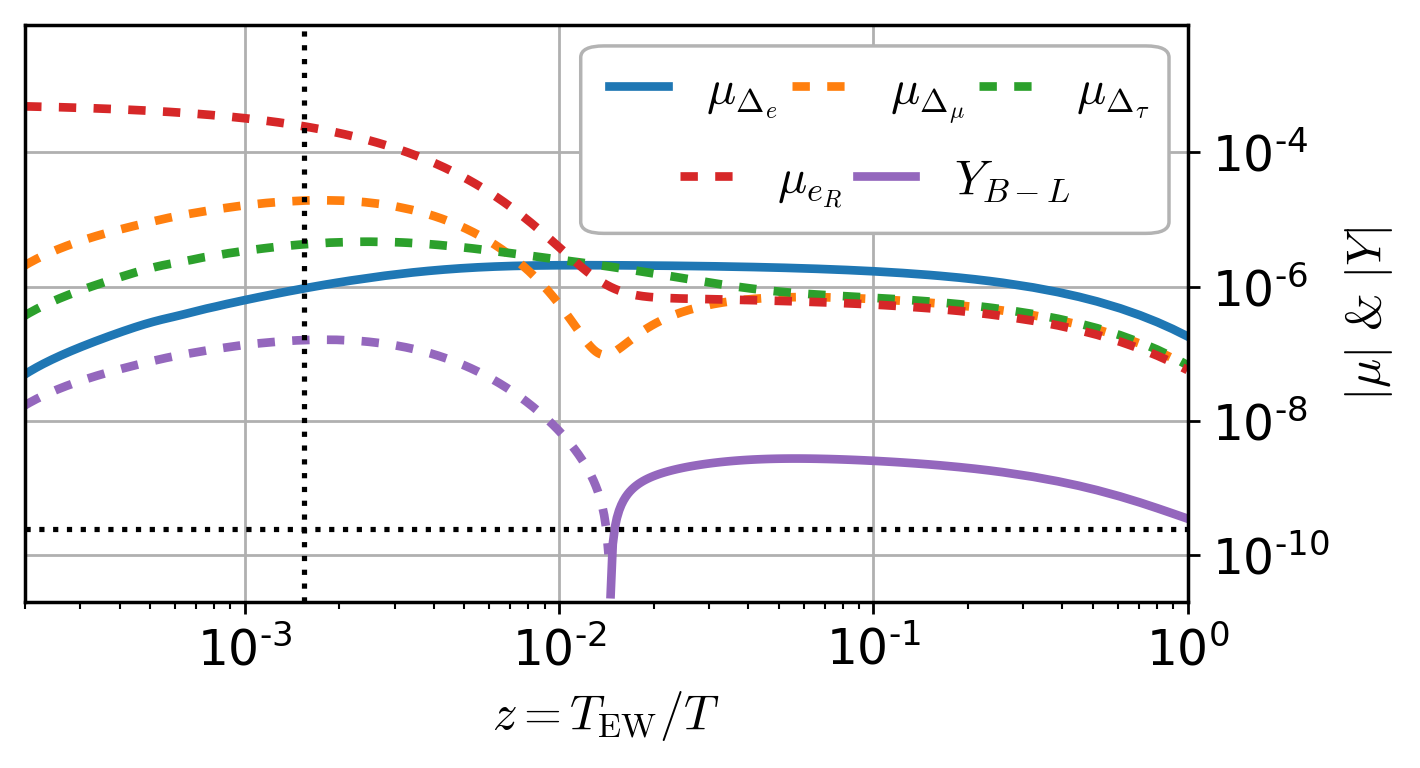} }}%
    \caption{Example solutions of the set of evolution equations in Eqs.~\eqref{eq:QKE1}-\eqref{eq:QKE2} and~\eqref{eq:beq_eR}. The left (right) panel was obtained with $M=1\, (1)$ GeV, $\delta M=0.5\, (0.5)$, ${\rm Re}\, \omega=\pi/4\, (\pi/4)$, ${\rm Im}\,\omega=1.6\, (5)$, and $\alpha_{31}=3\pi\, (3\pi)$. Solid\,/\,dashed lines denote positive\,/\,negative asymmetries. The vertical black-dotted curve indicates the right-handed electron equilibration temperature, and the horizontal dotted curve shows the value of $Y_{B-L}$ consistent with the observed BAU~\cite{Planck:2018vyg,ParticleDataGroup:2024cfk}. }
    \label{fig:examples}
\end{figure*}

\noindent\textbf{Numerical solutions}\,---\, We now proceed with a fully numerical analysis of the coupled system of equations in Eqs.~\eqref{eq:QKE1}-\eqref{eq:QKE2} and~\eqref{eq:beq_eR}~\cite{Ulysses}. 
For concreteness, we will henceforth consider the minimal type-I seesaw extension of the SM with two RHNs, which is the minimal setup consistent with the observed neutrino oscillations, while emphasizing that the WIFI-LG mechanism requires only a single RHN. Our analysis is carried out using the Casas-Ibarra parametrization~\cite{Casas:2001sr}, $F=\frac{1}{v}U_{\nu}\sqrt{m_\nu}R^T(\omega)\sqrt{M_N}$, with best-fit values for the Pontecorvo-Maki-Nakagawa-Sakata (PMNS) matrix, $U_\nu$, from Ref.~\cite{Esteban:2024eli}. There are five free parameters in the parametrization: Two RHN-mass parameters, $M_{1,2}\equiv M\left(1\pm\frac{\delta M}{2}\right)$, the real and imaginary components of the complex angle $\omega$ entering the orthogonal matrix $R$, and a Majorana phase $\alpha_{31}$ ($\alpha_{21}$) for normal (inverted) neutrino-mass ordering.

Two example solutions of Eqs.~\eqref{eq:QKE1}-\eqref{eq:QKE2} and~\eqref{eq:beq_eR} are shown in Fig.~\ref{fig:examples}, for a negligible initial RHN abundance and vanishing initial lepton-flavor asymmetries $R_N(T=\SI{e6}{\giga\electronvolt}) = \mu_{\alpha}(T=\SI{e6}{\giga\electronvolt}) = 0$, with CI parameters specified in the figure caption. Both panels were obtained using $\abs{\mu_{e_R}^0}=5\cdot 10^{-4}$. The panel to the left is consistent with the simple analytical reasoning presented above: After a period of freeze-in via wash-in, there remains a large hierarchy between $\mu_{e_R}$ and $\mu_{\Delta}$ at $T\sim T_e$, which prevents large subsequent wash-out of $\mu_\Delta$. In the panel to the right, the simple parametric estimate breaks down due to rapid equilibration of RHNs, making terms proportional to $\expval{\gamma_N^{(0)}}$ critical for a reliable description. In this case, the evolution of lepton-flavor asymmetries features enhanced wash-in compared to the first example, and also stronger subsequent wash-out.

\noindent\textbf{Parameter space}\,---\, It is useful to present the LG parameter space in terms of the active-sterile neutrino mixing and the masses of the RHNs. The neutrino flavor eigenstates can be written in terms of the three light and two heavy mass eigenstates, $\nu_i$ and $N_I$, as $\nu_{\alpha} = U_{\alpha i}\nu_i + \Theta_{\alpha I}N_I^c$, where $\Theta_{\alpha I}$ denotes the mixing between active neutrinos and the RHNs. A convenient measure of the overall mixing strength is $U^2 = \sum_{\alpha,I}\abs{\Theta_{\alpha I}}^2$. While the seesaw mechanism imposes a lower bound on this quantity, $U^2 \gtrsim \sum_i m_i / M$, the requirement of successful LG places an upper bound on $U^2$~\cite{Klaric:2020phc}.

\begin{figure}
    \centering
    \includegraphics[width=\linewidth]{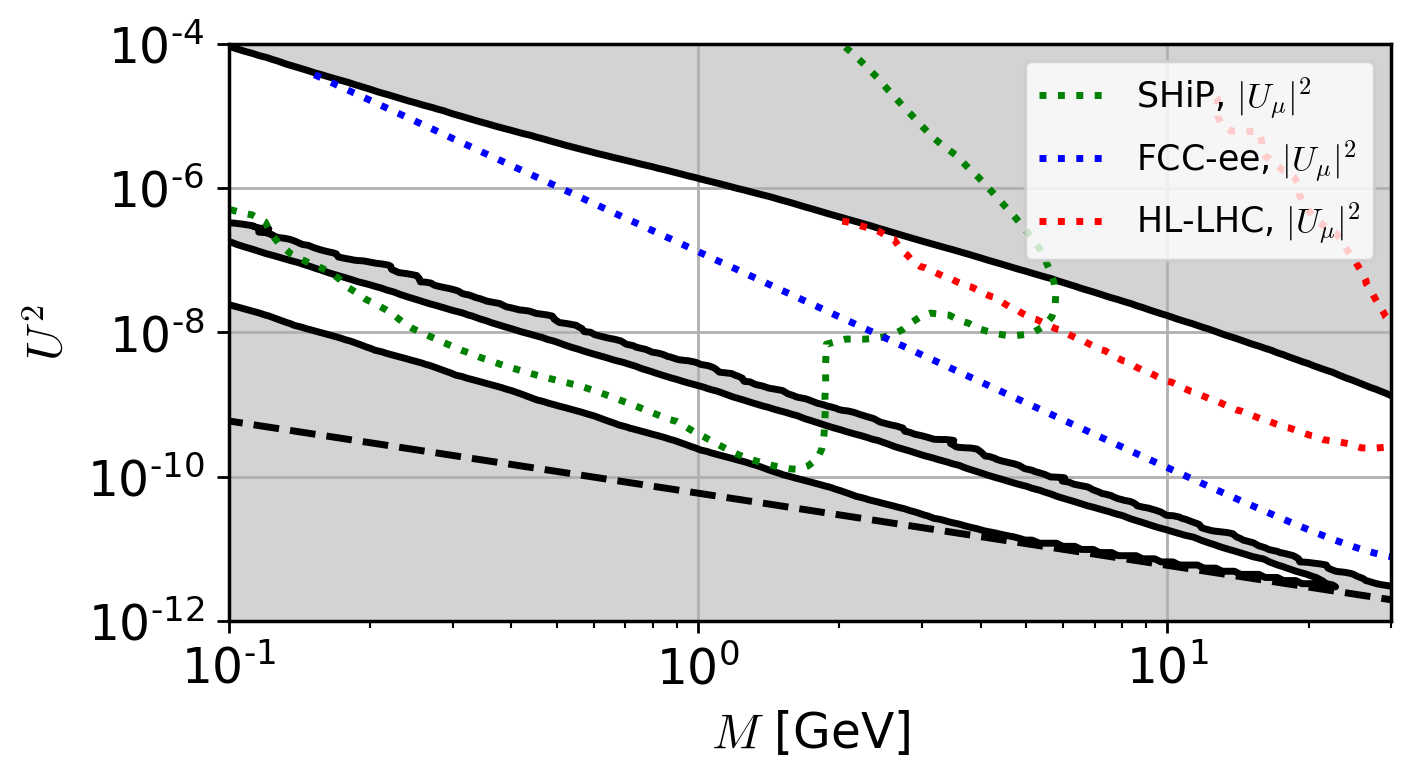}
    \caption{ Regions of parameter space where the BAU can be fully accounted for by WIFI-LG in the minimal type-I seesaw model for RHN-mass splitting $\delta M=0.5,{\rm Re}\, \omega=\pi/4$ and initial conditions respecting the bound in Eq.~\eqref{eq:upperbound}. The green, red, and blue curves display expected sensitivities of SHiP ($\lvert{U_{\mu}}\rvert^2$)~\cite{SHiP:2018xqw}, HL-LHC ($\lvert{U_{\mu}}\rvert^2$)~\cite{Drewes:2019fou}, and FCC-ee ($\lvert{U_{\mu}}\rvert^2$)~\cite{Drewes:2025ocf}, respectively. The experimental sensitivities presented here are intended merely as a qualitative guide, as they apply only for the mixing $\lvert{U_{\mu}}\rvert^2$, and the large mass splitting, $\delta M=0.5$, requires a more careful treatment of the associated constraints. See main text and Appendix~\ref{appendix:B} for further details. 
    }
    \label{fig:parameter_space}
\end{figure}

In Fig.~\ref{fig:parameter_space}, we show the region of parameter space, $(M,\,U^2)$, where the observed value of the BAU can be accounted for through WIFI-LG in the minimal type-I seesaw model for fixed RHN-mass splitting $\delta M=0.5$ and ${\rm Re}\, \omega=\pi/4$. The generated asymmetry depends only mildly on the fixed parameters, $\delta M$ and ${\rm Re}\,\omega$, and we scan over the remaining three free parameters. The parameter space is bounded from below by the seesaw line shown by the black dashed curve. While the parameter space can be further constrained by direct searches and BBN, these bounds depend on the relative mixing pattern $U_e^2:U_{\mu}^2:U_{\tau}^2$~\cite{Eijima:2018qke}. To not rule out parameter regions on the basis of these limits alone~\cite{deVries:2024rfh}, we leave a detailed study of the flavor dependence of the parameter space and a careful account of constraints for future work. However, note that RHN masses $M\lesssim10^{-1}$ GeV are severely constrained by BBN irrespective of the flavor structure and therefore not considered here. Moreover, for RHN masses heavier than roughly $40$ GeV, non-relativistic corrections to the rates~\cite{Hernandez:2022ivz,Granelli:2023vcm} as well as contributions from freeze-out~\cite{Klaric:2020phc} might become important requiring further investigation. The result shown in Fig.~\ref{fig:parameter_space} was obtained assuming normal mass hierarchy for active neutrinos and the corresponding result for inverted mass hierarchy is included in the appendix. To contextualize the figure, note that standard FILG without wash-in cannot generate the BAU in any region of parameter space once $\delta M \gtrsim 10^{-2}$, assuming a normal-ordered neutrino mass spectrum in the minimal type-I seesaw model~\cite{Klaric:2020phc}. 
Meanwhile, Fig.~\ref{fig:parameter_space} illustrates that through our mechanism, LG remains a viable explanation of the BAU even for GeV-scale RHNs with $\delta M\gg 10^{-2}$. Additional details on the visible features and a more comprehensive comparison between standard FILG and WIFI-LG can be found in the appendix. It is worth stressing that the WIFI-LG mechanism enables the type-I seesaw extension of the SM to account for the BAU through FILG for a mass spectrum featuring at least one RHN at the GeV scale but otherwise \textit{arbitrary} mass splittings, even in the minimal scenario where there are only two RHNs. Intriguingly, large regions of parameter space of the minimal seesaw model will be within reach of planned experiments, as indicated by the sensitivity curves in Fig.~\ref{fig:parameter_space}, making WIFI-LG a baryogenesis scenario that will be both testable -- and falsifiable -- with particle-physics experiments. Finally, our numerical analysis shows that WIFI-LG remains a viable explanation of the BAU for values of $\abs{\mu_{e_R}^0}$ as low as $6\%$ of the upper bound in Eq.~\eqref{eq:upperbound}.

\noindent \textbf{Conclusion and outlook}\,---\,In this letter, we have provided a framework where successful baryogenesis through FILG can be realized even with a single RHN around the GeV scale. As a proof-of-concept, we considered the minimal type-I seesaw extension of the SM with RHN masses $0.1\,\text{GeV}\lesssim M_N\lesssim 40$ GeV and demonstrated that the BAU can be successfully produced throughout wide regions of parameter space where the mass splitting between the two RHNs is too large for conventional FILG to be viable. This new freeze-in mechanism, which we term WIFI-LG, does not require sizable CP violation in the RHN sector and will be testable at planned intensity experiments and future colliders. 

Our work provides many opportunities for further exploration. Although we have focused on the freeze-in contribution for RHN masses $M_N \lesssim \SI{40}{\giga\electronvolt}$, the analysis could -- and should -- be extended to heavier RHN masses. Intriguingly, by employing the formalism outlined in \cite{Klaric:2020phc}, it should be possible to unite freeze-out WILG and freeze-in WILG; we leave a detailed study of this for future work~\cite{MojahedAndWeber}. As emphasized in Appendix~\ref{appendix:B}, another interesting direction for future work is the extension of WIFI-LG to the type-I seesaw model with three RHNs. Moreover, the results presented in this work can be refined by improving aspects of the evolution of $q_e$, and they motivate dedicated analyses of the chiral plasma at temperatures $T \sim T_e$, as discussed in Ref.~\cite{Mojahed:2025vgf}. Last but not least, our study motivates further exploration of possible mechanisms capable of producing suitable initial conditions for WILG~\cite{Domcke:2022kfs,Schmitz:2023pfy,Mukaida:2024eqi,Mojahed:2024yus}.


\medskip\noindent
\textit{Acknowledgments}\,---\, We are grateful to Kai Schmitz for many valuable discussions. We would also like to thank Marco Drewes and Kai Schmitz for helpful comments on the draft.
The work of M.\,A.\,M.\ and S.\,W.\ was supported by the Cluster of Excellence “Precision Physics, Fundamental Interactions, and Structure of Matter” (PRISMA$^+$ EXC 2118/1) funded by the Deutsche Forschungsgemeinschaft (DFG, German Research Foundation) within the German Excellence Strategy (Project No. 390831469). 

\small

\bibliographystyle{JHEP} 
\bibliography{refs}

\begin{thebibliography}{10}
\bibitem{LNVcomment}
{The terms featuring lepton-number-violating interactions in Eqs.~\eqref{eq:QKE1}-\eqref{eq:QKE2} are parametrically suppressed by a factor of $M^2/T^2$ compared to the terms proportional to the LNC rates $\expval{\gamma^{(i)}}$.}
\bibitem{Osccomment}
{Recall that oscillations are imperative for asymmetry generation in standard FILG, hence neglecting oscillations provides a clear route to studying asymmetry-generation effects stemming solely from $\mu_{e_R}\neq 0$.}
\end{thebibliography}

\newpage
\onecolumngrid
\newpage


\appendix

\section{Inverted neutrino mass hierarchy}
\begin{figure}
    \centering
    \includegraphics[width=0.45\linewidth]{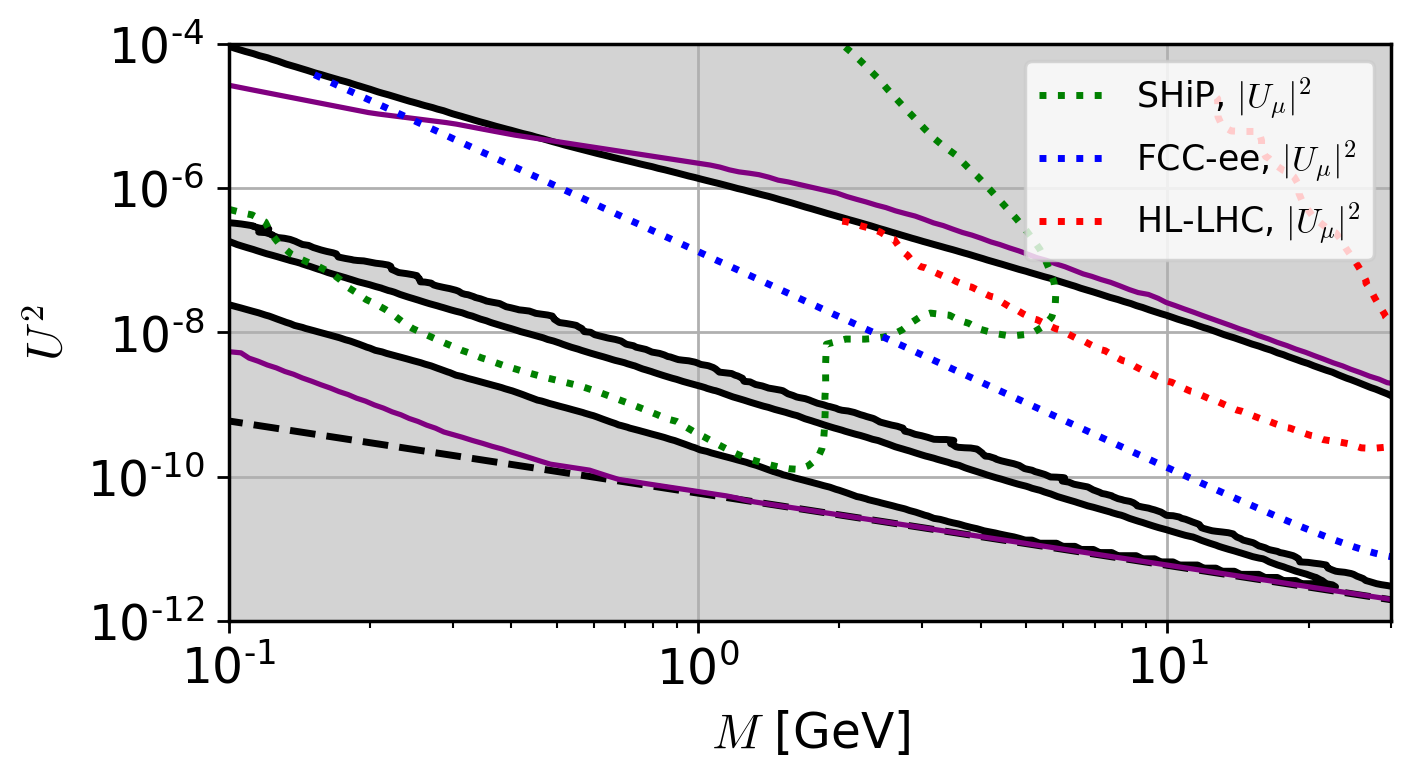}\quad 
    \includegraphics[width=0.45\linewidth]{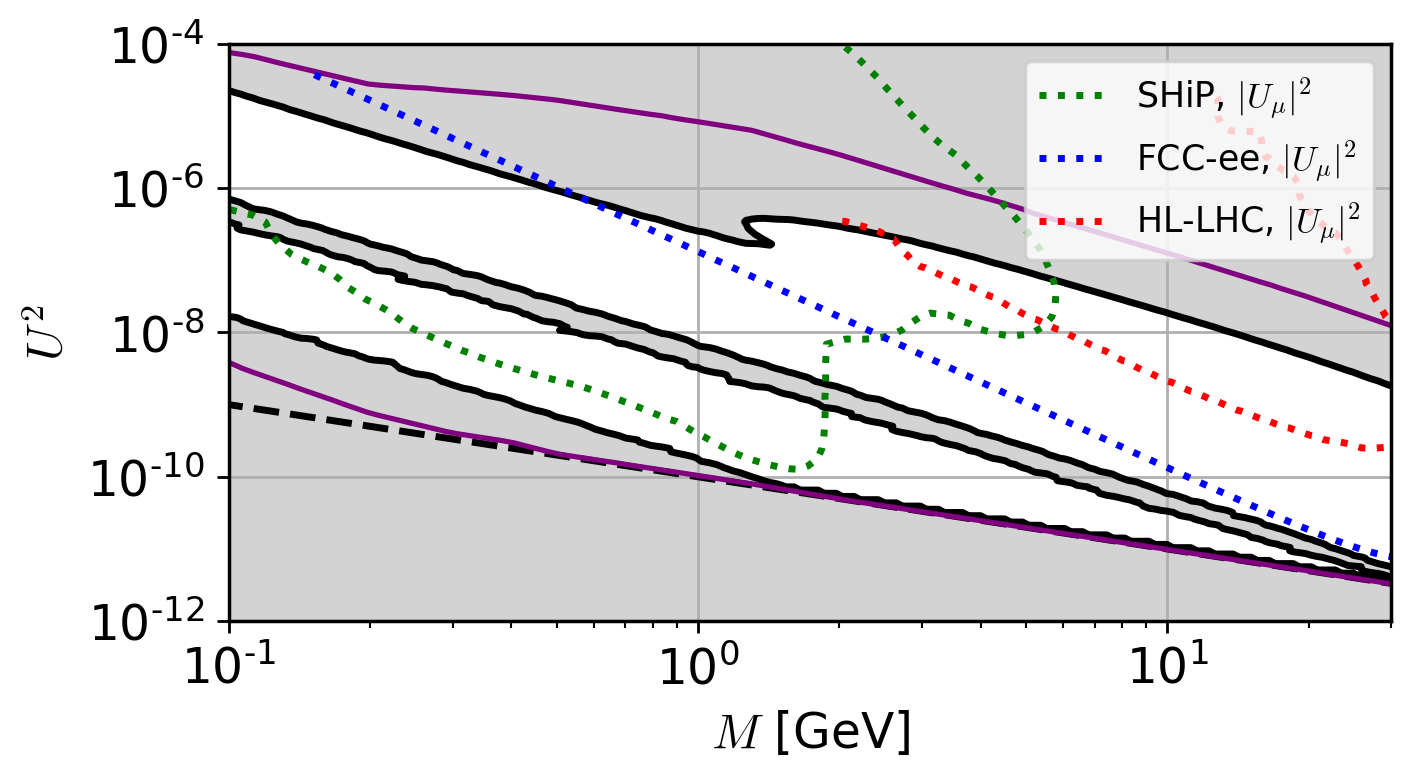}
    \caption{Regions of parameter space where the BAU can be fully accounted for by WIFI-LG in the minimal type-I seesaw model for RHN-mass splitting $\delta M=0.5,$ ${\rm Re} \, \omega=\pi/4$ and initial conditions respecting the bound in Eq.~\eqref{eq:upperbound} for normal (left) and inverted (right) neutrino-mass ordering. Allowing $\delta M$ to vary freely, the observed BAU can be explained through pure FILG in the parameter region enclosed by the purple curves, taken from Ref.~\cite{Klaric:2020phc}. 
    The green, red, and blue curves display expected sensitivities of SHiP ($\lvert{U_{\mu}}\rvert^2$)~\cite{SHiP:2018xqw}, HL-LHC ($\lvert{U_{\mu}}\rvert^2$)~\cite{Drewes:2019fou}, and FCC-ee ($\lvert{U_{\mu}}\rvert^2$)~\cite{Drewes:2025ocf}, respectively.
    }
    \label{fig:parameter_space_comparison}
\end{figure}

Throughout the main text, we considered normal mass ordering for neutrinos. In the right panel of Fig.~\ref{fig:parameter_space_comparison}, we show regions (in white) of parameter space in the minimal type-I seesaw model that are compatible with explaining both the BAU through WIFI-LG and neutrino masses through the seesaw mechanism for inverted neutrino mass ordering. The figure shows that both normal and inverted mass orderings admit comparatively large regions of parameter space compatible with successful WIFI-LG, and that large regions of these spaces can be tested by upcoming experiments. For comparison, we have included purple curves bounding the parameter space where standard FILG can account for the BAU when $\delta M$ is treated as a free parameter~\cite{Klaric:2020phc}.

\section{Comparison between standard FILG and WIFI-LG}
\label{appendix:B}

\noindent\textbf{Overview}\,---\, 
In the following, we elaborate on similarities and differences between standard FILG and WIFI-LG. To do so, it is useful to first recall the differences in how Sakharov's conditions are satisfied in WILG and standard LG in the type-I seesaw paradigm. Standard LG is based on the following two assumptions: 
\begin{itemize}
    \item RHNs play a fundamental role in satisfying all three of Sakharov's conditions. In particular, the required source of CP violation originates from the RHN sector. A lepton asymmetry is generated through CP-violating, lepton-number-violating RHN interactions that take place out of thermal equilibrium. Part of this lepton asymmetry is converted to a baryon asymmetry by efficient $B+L$-violating electroweak sphaleron interactions in the thermal plasma. 
    \item There are no asymmetries in any (approximately) conserved charges other than $q_{\Delta}$.
\end{itemize}
This should be contrasted with WILG, which in its purest form makes the following assumption: 
\begin{itemize}
    \item The CP violation required to satisfy Sakharov's conditions is not supplied in sufficient amount from the RHN sector. Instead, a sufficient amount of CP violation is provided from another charge-generation mechanism, referred to as chargegenesis, which takes place at a temperature scale that is above the mass scale set by the lightest RHN(s). The chargegenesis mechanism, which does not need to violate $B-L$, generates asymmetries in at least one of the global charges that are conserved in the SM at high temperatures. The role of the RHNs is then to provide a source of $B-L$ violation, which can drive the non-trivial chemical background of the thermal plasma to a state where $q_\Delta\neq 0$. Note that RHNs in WILG play a role in baryogenesis similar to that of electroweak sphalerons in standard LG. 
\end{itemize}
Both standard LG and WILG can be further divided into a freeze-in and a freeze-out regime, as explained in the main text. Our discussion therefore suggests that the landscape of LG mechanisms based on the type-I seesaw model can be characterized (at least schematically) by the following diagram:
\begin{center}
\begin{tikzpicture}[>=stealth, scale=1.2]

\draw[->] (0,-0.3) -- (6,-0.3) node[right] {$M_N$};
\draw[->] (-0.3,0) -- (-0.3,2) node[above] {$\mu^0$};

\draw[->] (-0.6,2) -- (-0.6,0) node[below] {$CPV_N$};

\draw (0,0) rectangle (6,2);

\draw (3,0) -- (3,2);

\draw (0,1) -- (6,1);

\node at (1.5,1.5) {Freeze-in WILG};
\node at (4.5,1.5) {Freeze-out WILG};
\node at (1.5,0.5) {Standard FILG};
\node at (4.5,0.5) {Standard FOLG};

\end{tikzpicture}
\end{center}
where $CPV_N$ denotes $CP$-violation originating from the RHN sector. 
The present work is confined to the (upper-)left part of this figure.
Of course, in reality, baryogenesis may be a mixture of freeze-in and freeze-out~\cite{Klaric:2020phc} and of wash-in and standard LG~\cite{Domcke:2020quw}. In particular, a scenario with substantial CP-violation in the RHN sector \text{and} a non-negligible background of primordial charge-asymmetries would blur the line between the upper - pure WILG - part of the figure, and the lower - standard LG - part of the figure. A description that unifies all of the boxes above remains an important task for the future. In this appendix, we will focus on LG via freeze-in, i.e. the left-part of the figure, and consider normal neutrino-mass hierarchy for concreteness.

\begin{figure*}[t]%
    \centering
    {{\includegraphics[height=0.19\textheight]{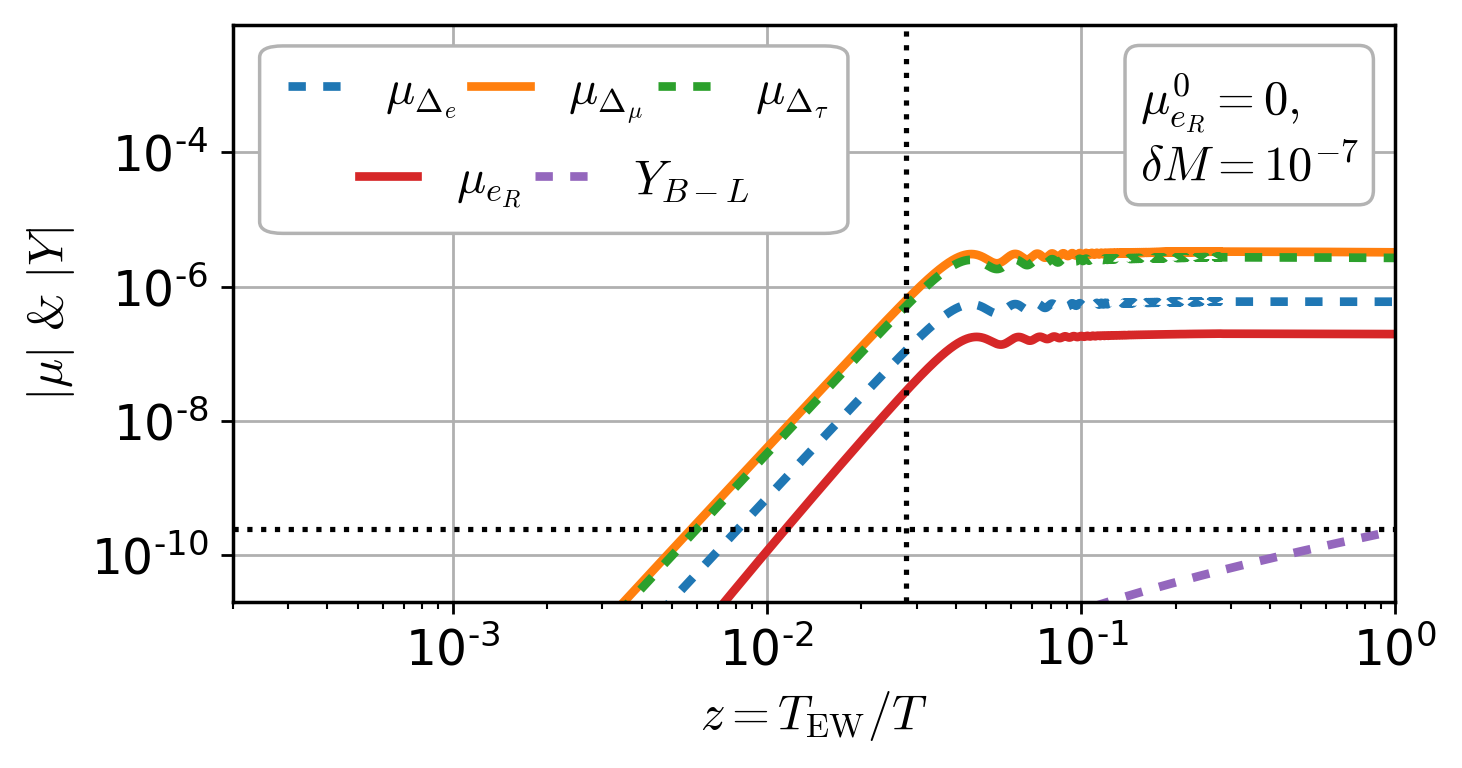}}}%
    \quad
    {{\includegraphics[height=0.19\textheight]{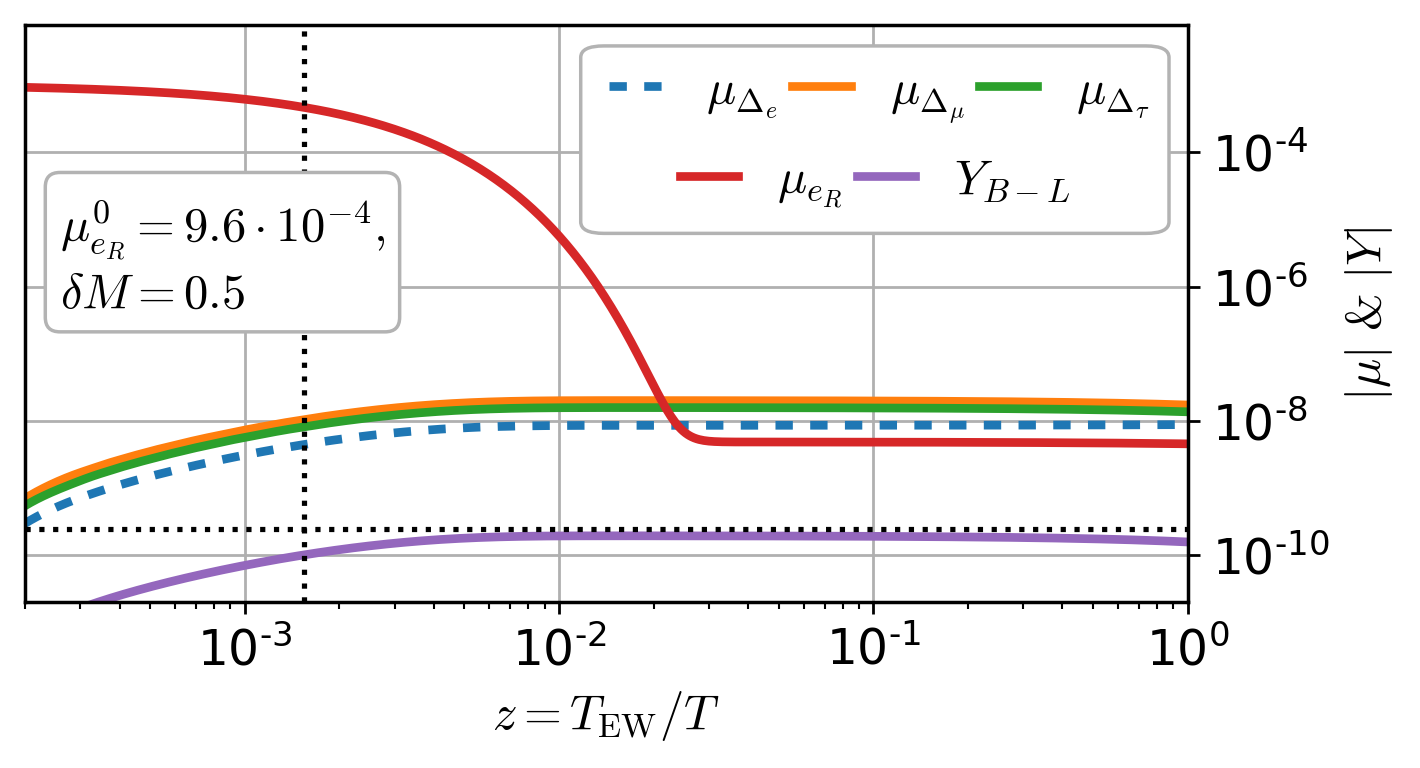} }}
    {{\includegraphics[height=0.19\textheight]{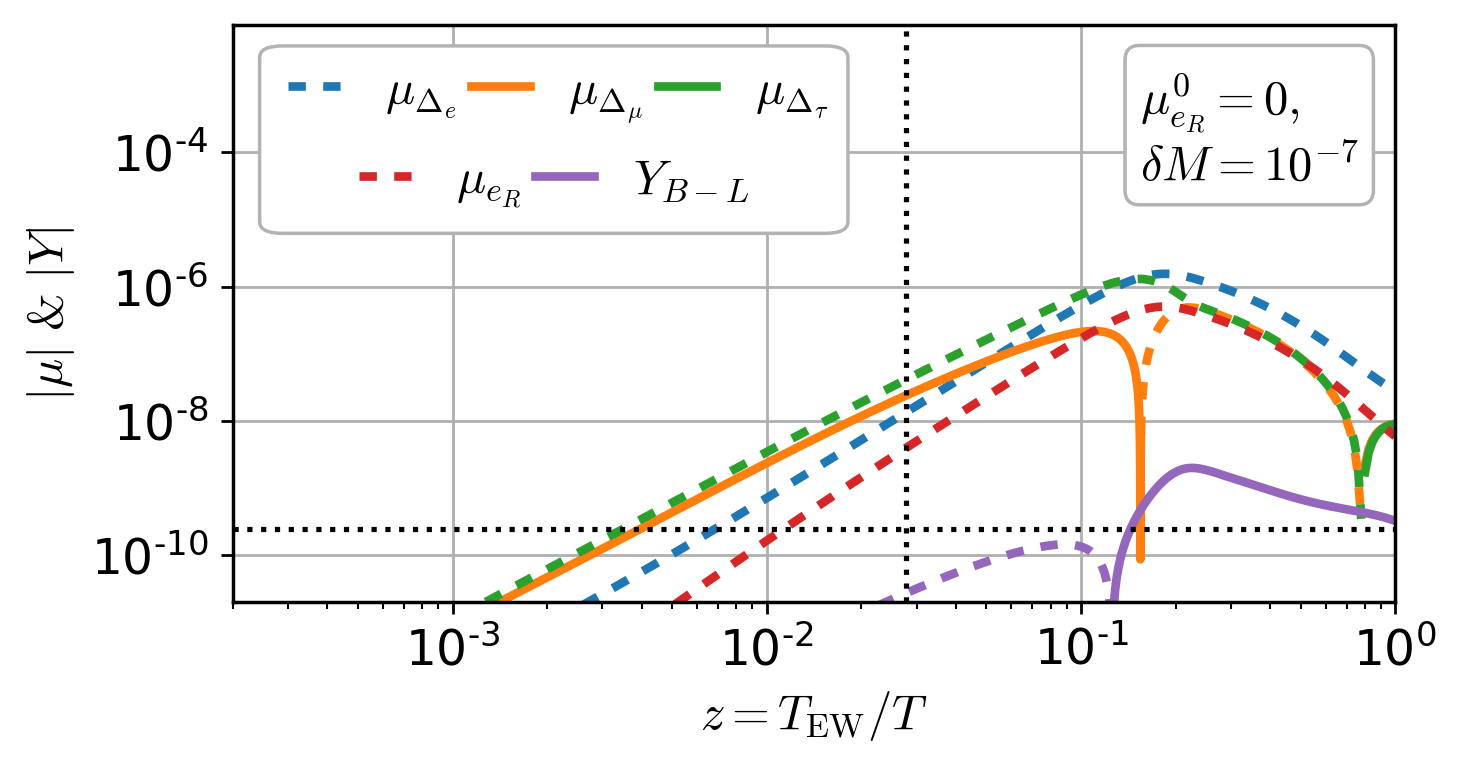}}}%
    \quad
    {{\includegraphics[height=0.19\textheight]{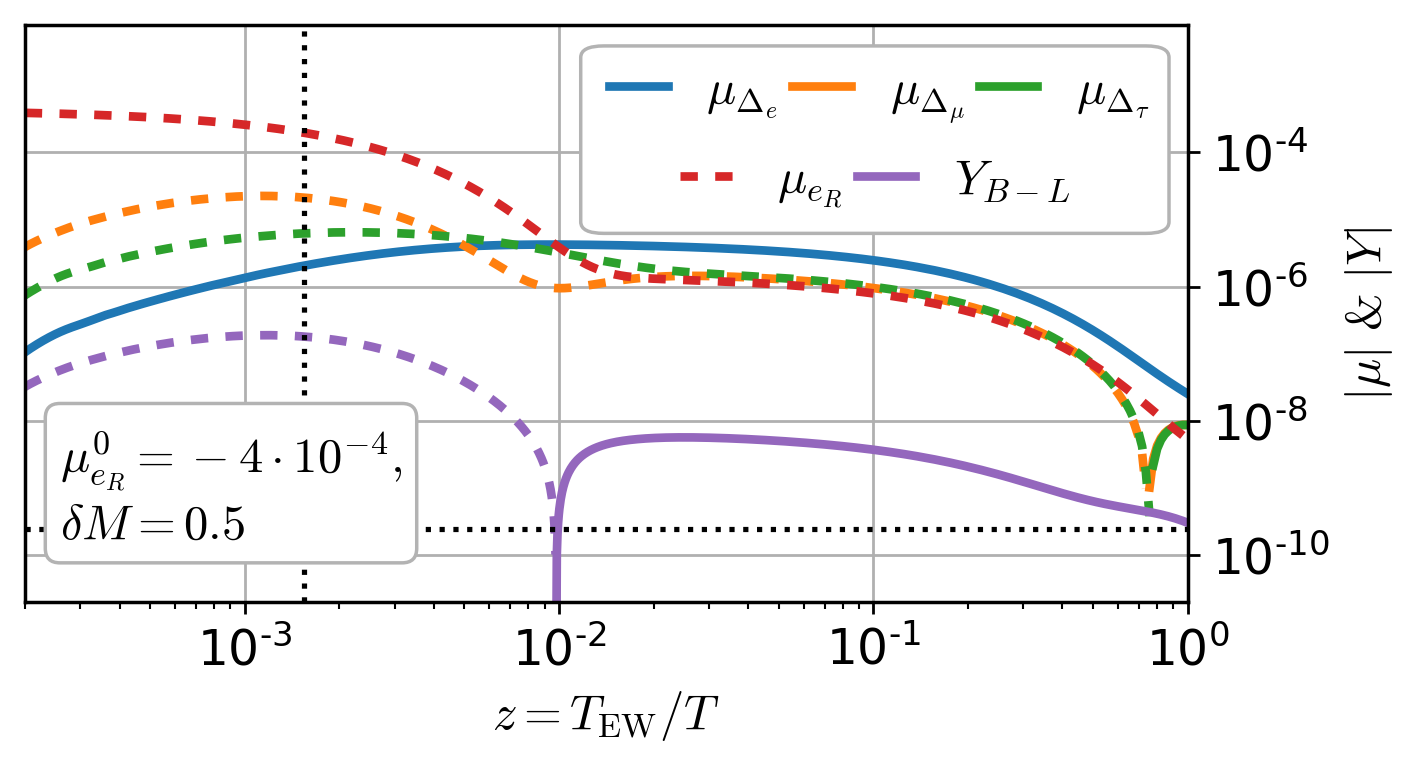} }}%
    \caption{Time evolution of asymmetries for pure FILG (left) and pure WIFI-LG (right). The values for $\delta M$ and $\mu_{e_R}^0$ are indicated in the panels, and the remaining neutrino parameters are fixed as $M=\SI{3}{\giga\electronvolt},$ $\alpha_{31}=3\pi$, ${\rm Re}\,\omega=\pi/4$, and $ {\rm Im}\, \omega=0\  (4.9)$ in the top (bottom) panels.}
    \label{fig:comparison}
\end{figure*}

\noindent\textbf{Time evolution}\,---\, It is instructive to compare time-evolution of asymmetries in WIFI-LG with standard FILG. In WIFI-LG, the generation of lepton-flavor asymmetries is a result of nonzero $\mu^0$ and RHN interactions. The generation of $q_\Delta$ is efficient while $\mu^0$ is sizable, so the characteristic temperature scale associated with asymmetry generation is set by $T_e$. Meanwhile, in standard FILG, the generation of asymmetry is driven by CP-violating coherent oscillations of relativistic RHNs. Asymmetry generation may become efficient around the characteristic temperature scale $T_{\rm{osc}}= \left(\Delta M_N^2M_0/12\right)^{1/3}$~\cite{Shuve:2014zua,Granelli:2023vcm}, which parametrizes the time when the oscillation rate between RHNs is similar to the Hubble expansion~\cite{Drewes:2017zyw}. The different mechanisms are illustrated in Fig.~\ref{fig:comparison}, which shows time-evolution of asymmetries in standard FILG (left) and WIFI-LG (right). In the right panels, the mass splitting $\delta M$ is sufficiently large that the asymmetry generated through oscillations is always negligible compared to the observed BAU in the minimal type-I seesaw model. Meanwhile, in the left panels we set $\mu_{e_R}^0=0$ and consider a small mass splitting. The upper row corresponds to small active-sterile mixing (on the seesaw line) and the lower row to large active-sterile mixing, see the figure and its caption for precise values for all neutrino parameters and $\mu_{e_R}^0$. In the standard FILG scenarios, the upper-left panel corresponds to the oscillatory regime, and the lower-left panel to the overdamped regime~\cite{Drewes:2016gmt}. The vertical black-dotted curve shows $T_{\text{osc}}$ for the standard FILG scenarios and $T_e$ for the WIFI-LG scenarios. The figure clearly illustrates that asymmetry generation proceeds through different production mechanisms in WIFI-LG and standard FILG, and that the associated characteristic temperature scales also differ.  

\begin{figure}
    \centering
    \includegraphics[width=0.45\linewidth]{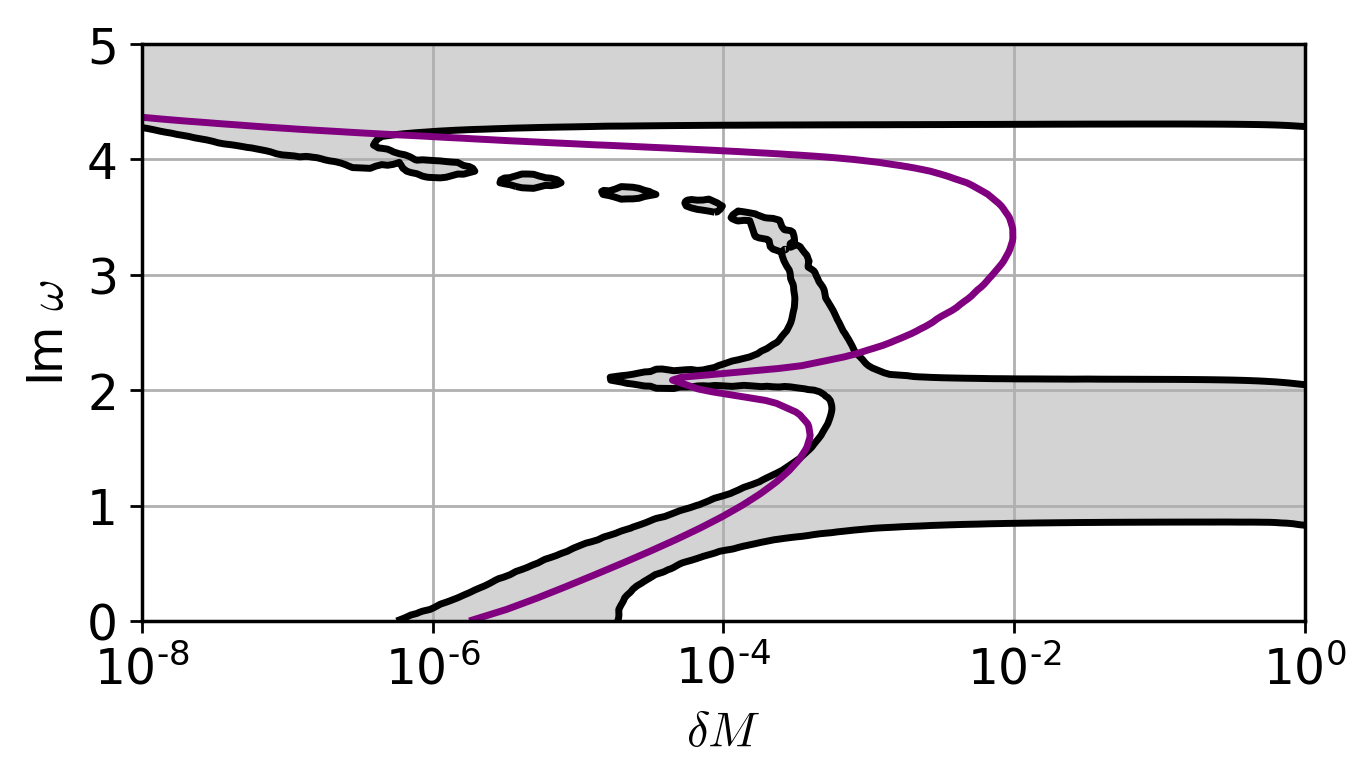} \quad
    \includegraphics[width=0.45\linewidth]{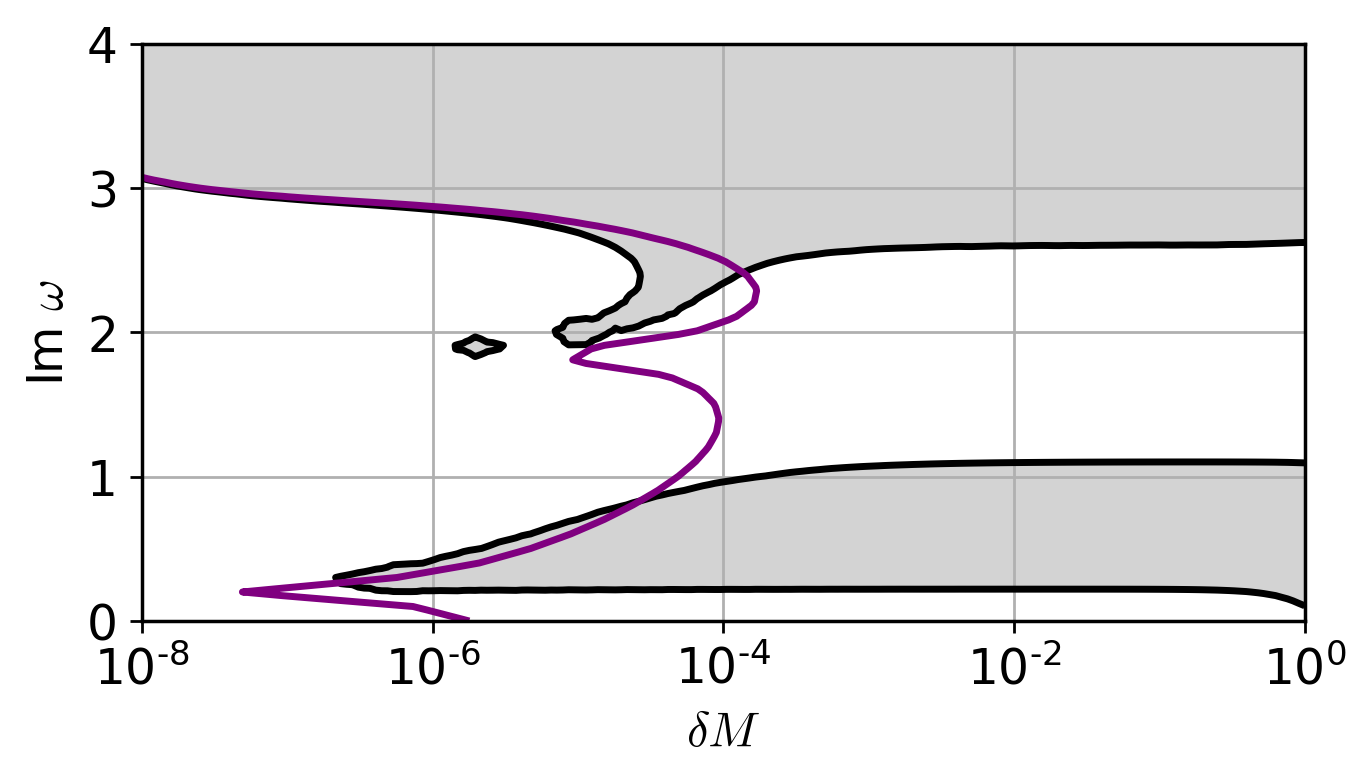}
    \caption{White: Region of $\delta M$-Im$\omega$ plane where the BAU can be fully accounted for by WIFI-LG in the minimal type-I seesaw model with normal neutrino-mass ordering for $M=\SI{10}{\giga\electronvolt}$, ${\rm Re}\, \omega = \pi/4$, and $\alpha_{31}=-\pi\,(+\pi)$ left (right). The purple curve shows the region where pure FILG can account for the BAU.
    }
    \label{fig:mass_splitting}
\end{figure}

\noindent\textbf{Parameter space}\,---\, Finally, let us compare available parameter space for standard FILG and WIFI-LG in the minimal type-I seesaw model. The white region in the panel to the left in Fig.~\ref{fig:parameter_space_comparison}, which was also shown in Fig.~\ref{fig:parameter_space} in the main text, is the available parameter space for WIFI-LG for $\delta M=0.5$. Due to the interplay of wash-in, wash-out, and flavor effects, the parameter space is separated into two disjoined parts by a thin gray region in Fig.~\ref{fig:parameter_space_comparison}. However, it should be noted that the maximum baryon asymmetry attainable in this region satisfies $Y_B>Y_{B}^{\rm{obs}}/2$, so we expect it could be significantly deformed in future precision studies, for example by explicitly studying the momentum dependence of the QKEs which can change the baryon asymmetry by $\order{1}$~\cite{Asaka:2011wq,Ghiglieri:2018wbs}. 
The region between the purple curves is the parameter space where standard FILG can account for the BAU when $\delta M$ is treated as a free parameter. Although the region bounded by the purple curve largely overlaps with the white region, there are regions of parameter space that are compatible with only one of the LG mechanisms, where a discovery would immediately favor one of the two. In the scenario of a discovery where the two regions overlap, additional information about the RHN-mass splitting could be decisive in determining the underlying LG mechanism. This point is illustrated for normal ordering in Fig.~\ref{fig:mass_splitting}, which shows the region in the $\delta M$-${\rm Im}\,\omega$ plane for $M=10$ GeV where the two mechanisms can account for the BAU. In the left panel, the remaining free neutrino parameters, ${\rm Re}\,\omega$ and $\alpha_{31}$, were fixed to values favorable for standard FILG, and we show an alternative choice for comparison in the right panel. The purple curve in the left panel is consistent with Fig.~10 in \cite{Klaric:2021cpi}. The important point is that for sufficiently large mass splittings, FILG cannot account for the BAU in the minimal type-I seesaw model, and WIFI-LG becomes the only viable candidate. Hence, a discovery in the white region of Fig.~\ref{fig:parameter_space_comparison} could favor WIFI-LG if it is accompanied by an experimental determination of $\delta M$ indicating mass splittings at the percent level or larger. 

Let us finally emphasize that in this letter we have restricted ourselves to the minimal type-I seesaw framework with two RHNs. It has long been appreciated in the FILG literature~\cite{Drewes:2012ma} that the inclusion of a third RHN can qualitatively alter the dynamics of asymmetry generation~\cite{Abada:2018oly} and substantially enlarge the region of parameter space compatible with successful baryogenesis via LG~\cite{Drewes:2021nqr}. In light of these developments in standard FILG, we anticipate that extending WIFI-LG to the type-I seesaw model with three RHNs may similarly reveal novel mechanisms for asymmetry generation and further broaden the viable parameter space. We therefore leave a systematic exploration of this scenario for future work.

\end{document}